\documentclass[11pt]{article}

\usepackage[preprint]{acl}

\usepackage{times}
\usepackage{latexsym}

\usepackage[T1]{fontenc}

\usepackage[utf8]{inputenc}

\usepackage{microtype}

\usepackage{inconsolata}

\usepackage{graphicx}
\usepackage{subcaption}
\usepackage{hyperref}
\usepackage{amsmath,amssymb,amsfonts,graphicx}
\usepackage{color}
\usepackage{colortbl}
\usepackage{textcomp}
\usepackage{xcolor}
\usepackage{booktabs}
\usepackage{makecell}
\usepackage{array}
\usepackage[table]{xcolor}
\usepackage{multirow,multicol}
\usepackage{graphicx}
\usepackage{tabularx}
\usepackage{diagbox}
\usepackage{enumitem}
\usepackage{adjustbox}
\usepackage{pifont}
\usepackage{cmtt}
\usepackage{float}
\usepackage{stfloats}
\usepackage{algorithm,algpseudocode}
\usepackage[flushleft]{threeparttable}
\usepackage{url}
\newcolumntype{Y}{>{\centering\arraybackslash}X}
\newcolumntype{C}[1]{>{\centering\arraybackslash}p{#1}}
\definecolor{blue800}{HTML}{1565C0}
\definecolor{pink800}{HTML}{AD1457}
\definecolor{blue100}{HTML}{BBDEFB}
\definecolor{blue50}{HTML}{E3F2FD}
\definecolor{indigo50}{HTML}{E8EAF6}
\definecolor{indigo700}{HTML}{303F9F}
\definecolor{green50}{HTML}{E8F5E9}
\definecolor{teal50}{HTML}{E0F2F1}
\definecolor{red50}{HTML}{FFEBEE}
\definecolor{purple50}{HTML}{EDE7F6}
\definecolor{grey100}{HTML}{eaebef}
\definecolor{grey300}{HTML}{d0d1db}
\definecolor{hyperlink}{HTML}{000099}
\hypersetup{
    colorlinks=true,
    linkcolor=hyperlink,
    citecolor=hyperlink,
    filecolor=hyperlink,
    urlcolor=hyperlink
}
\title{A Data-Centric Approach to Generalizable Speech Deepfake Detection}

\author{
 \textbf{Wen Huang\textsuperscript{$\sharp$, $\flat$}},
 \textbf{Yuchen Mao\textsuperscript{$\sharp$, $\flat$}},
 \textbf{Yanmin Qian\textsuperscript{$\sharp$, $\flat$}\thanks{ Corresponding author}}
\\
\\
\textsuperscript{$\sharp$}Auditory Cognition and Computational Acoustics Lab\\
MoE Key Lab of Artificial Intelligence, AI Institute\\
School of Computer Science, Shanghai Jiao Tong University, China\\
\textsuperscript{$\flat$}VUI Labs, China
}

\begin{document}
\maketitle
\begin{abstract}
Achieving robust generalization in speech deepfake detection (SDD) remains a primary challenge, as models often fail to detect unseen forgery methods. While research has focused on model-centric and algorithm-centric solutions, the impact of data composition is often underexplored. This paper proposes a data-centric approach, analyzing the SDD data landscape from two practical perspectives: constructing a single dataset and aggregating multiple datasets. To address the first perspective, we conduct a large-scale empirical study to characterize the data scaling laws for SDD, quantifying the impact of source and generator diversity. To address the second, we propose the Diversity-Optimized Sampling Strategy (DOSS), a principled framework for mixing heterogeneous data with two implementations: DOSS-Select (pruning) and DOSS-Weight (re-weighting). Our experiments show that DOSS-Select outperforms the naive aggregation baseline while using only 3\% of the total available data. Furthermore, our final model, trained on a 12k-hour curated data pool using the optimal DOSS-Weight strategy, achieves state-of-the-art performance, outperforming large-scale baselines with greater data and model efficiency on both public benchmarks and a new challenge set of various commercial APIs.
\end{abstract}

\section{Introduction}

\begin{figure}[t]
    \centering
    \includegraphics[width=\linewidth, trim={0mm, 1mm, 0mm, 0mm}
    ]{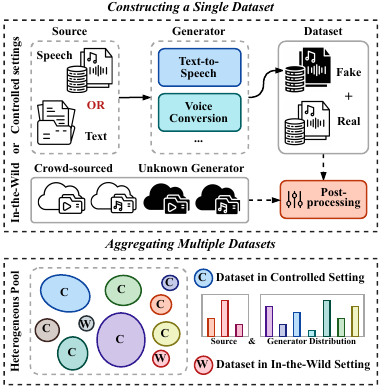}
    \caption{\textbf{The data landscape for speech deepfake detection}, from constructing a single dataset to aggregating multiple heterogeneous domains.}
    \label{fig:intro}
\end{figure}


Speech Deepfake Detection (SDD) has emerged as a critical research area in recent years. As speech synthesis technologies become increasingly sophisticated, the resulting deepfakes are often indistinguishable from authentic speech to the human ear, which poses significant security risks. This rapid progress introduces the primary unsolved challenge for detection: generalization. Detectors frequently fail when confronted with unseen forgery methods from ever-evolving generative systems. Furthermore, their performance can degrade significantly due to shifts in acoustic conditions, such as channel effects or background noise. This creates a critical need for detectors that are not only accurate but also robustly generalizable.

The community has largely pursued model-centric and algorithm-centric solutions to improve generalization. These include exploring novel architectures~\cite{jung2022aasist}, leveraging large self-supervised (SSL) pretrained models~\cite{tak2022automatic}, and adopting advanced training algorithms~\cite{huang2025from} or data augmentation techniques~\cite{tak2022rawboost}. A common limitation of these works, however, is their reliance on fixed and limited training benchmarks. In parallel, a diverse array of new datasets for SDD has emerged~\cite{zhao2024emofake, bhagtani2025diffssd}; yet, these valuable resources are typically used in isolated train/test settings or are simply held out as unseen test sets. We argue that a crucial yet often underexplored factor is data composition.

This paper aims to achieve generalizable SDD through a data-centric perspective. Unlike traditional audio tasks that rely on collected real-world data, the main portion of data for SDD is synthetically generated. This provides a unique opportunity to control and systematically study the data composition process, as a theoretically infinite number of samples can be generated. 

To formalize our study, we analyze the SDD data landscape from two practical perspectives, as illustrated in Figure~\ref{fig:intro}. First, we examine the process of constructing a single dataset. This can occur in two primary settings. In a controlled setting, developers typically begin with a known source (often real speech datasets that provide speech or text) and then select a specific generator (e.g., text-to-speech or voice conversion) to synthesize deepfakes. In an in-the-wild setting, audio is often crowdsourced from online platforms, where the underlying source and generator are typically unknown. Despite these differences in methodology, we identify two fundamental components that define any synthetic sample: the \textit{source} it was derived from and the \textit{generator} used to create it.
Second, we address the common task of aggregating multiple datasets. Given that each dataset possesses a distinct combination of sources, generators, and acoustic conditions, this aggregation inevitably creates a heterogeneous data pool. This resulting pool is characterized by significant variance in data size, source types, and synthesis methods, making it a complex target for training. This framework motivates two central research questions:
\begin{itemize}[leftmargin=*, itemsep=0pt]
    \item \textbf{RQ1:} When constructing a dataset, what principles determine how to allocate resources between source/generator diversity and data volume?
    \item \textbf{RQ2:} When aggregating multiple datasets, what is the most effective and efficient data mixing strategy to maximize generalization?
\end{itemize}

Our investigation into these questions leads to the following key contributions:
\begin{itemize}[leftmargin=*, itemsep=0pt]
    \item \textbf{Discovering Data Scaling Laws.} We conduct the first large-scale empirical study to characterize the data scaling laws for SDD, quantifying the distinct impacts of source and generator diversity on model generalization (Section~\ref{sec:scale}).
    \item \textbf{A Principled Data Mixing Strategy.} Based on these scaling laws, we propose the Diversity-Optimized Sampling Strategy (DOSS), a principled framework for efficiently training on heterogeneous data mixtures (Section~\ref{sec:doss}).
    \item \textbf{State-of-the-Art Generalization.} We validate our data-centric approach by applying the DOSS framework to a large-scale, heterogeneous data pool, achieving state-of-the-art generalization performance (Section~\ref{sec:final}).
\end{itemize}

\section{Related Work}
\paragraph{Generalization in SDD.}
Generalization remains a core challenge in SDD. A large proportion of research has sought to improve this by focusing on model design. Initial attempts included specialized, compact models designed to effectively extract temporal and spectral information from audio~\cite{tak2021end,jung2022aasist}. More recently, self-supervised learning (SSL) pre-trained models have become a common approach. SSL front-ends like Wav2Vec2~\cite{baevski2020wav2vec}, XLS-R~\cite{babu2021xlsr}, and WavLM~\cite{chen2022wavlm} are used to extract rich, intrinsic speech representations, which are then fine-tuned with a classifier backend. This method has shown strong performance and improved generalization~\cite{tak2022automatic,guo2024audio}. A parallel direction focuses on algorithmic improvements to the training strategy. This includes advanced data augmentation techniques~\cite{tak2022rawboost,wang2024can}, novel training objectives~\cite{zhang2021one,huang2025generalizable}, and refined optimization processes~\cite{huang2025from}.

While these attempts have advanced the field, most rely on fixed and limited training benchmarks, largely overlooking the impact of data composition. A few recent exceptions have begun to explore a data-centric paradigm to address these limitations. For instance, \citealp{combei2025unmasking} bridge the gap between scientific and real-world deepfakes using dataset selection and sample-level pruning, such as margin-based selection, to remove redundant or noisy data. In parallel, the work by \citealp{antideepfake_2025} demonstrated that post-training SSL models on 74k hours of speech improve generalization to unseen deepfakes; yet, this result relied on naive data aggregation which combined sources without regarding their variations. This paper builds upon this line of inquiry, shifting the focus from the sheer volume of data to its principled composition.

\paragraph{Scaling Laws.} Recent research has found that neural model performance scales predictably as a power-law of data size, model size, or computation~\cite{kaplan2020scaling}. This principle has been shown to apply broadly across diverse domains, including computer vision~\cite{zhai2022scaling}, multi-modal learning~\cite{aghajanyan2023scaling}, robotic manipulation~\cite{lin2025data}, and speech recognition~\cite{chen2025owls}. Understanding these laws is crucial for informing training decisions~\cite{hoffmann2022training} and enabling more effective resource allocation during model development~\cite{achiam2023gpt}. In this work, we extend this paradigm to SDD, examining how generalization performance scales with source and generator diversity, as well as sample volume, to inform principled data collection strategies.

\paragraph{Data Mixing.}
A highly relevant line of research is data mixing, which explores how to combine data from different sources for training large-scale models. Early approaches relied on manual heuristics, such as applying sampling caps to prevent domain dominance~\cite{raffel2020exploring} or explicitly oversampling domains perceived to be of high quality~\cite{brown2020language}. More recent work has automated this process. These methods are typically either offline, using a proxy model to find a static set of optimal weights before training~\cite{xie2023doremi,fan2024doge,liu2025regmix}, or online, dynamically adjusting sampling weights during training based on the model's real-time state~\cite{albalak2023efficient,chen2023skill}.
A key limitation of many of these automated methods is their reliance on optimizing performance for a fixed set of known domains. We argue that performance on a specific validation set does not necessarily reflect true generalization to completely unseen data. In contrast, our approach avoids proxy models and their reliance on fixed validation sets. We instead use a more fine-grained domain definition and a sampling strategy derived from fundamental scaling principles, making it inherently designed for robust out-of-domain generalization.

\section{Discovering Data Scaling Laws}

\label{sec:scale}
To answer RQ1, this section aims to empirically discover the data scaling laws for SDD by investigating two fundamental factors: \textit{generator diversity} and \textit{source diversity}. Our objective is to model the relationship between generalization performance and these key data composition variables:
\begin{itemize}[leftmargin=*, itemsep=0pt]
    \item $N_S$: The number of disctinct sources.
    \item $N_G$: The number of distinct generators.
    \item $V$: The volume of samples per unit of diversity.
\end{itemize}

\subsection{Experimental Setup}
To isolate the effects of data composition, all scaling law experiments share a common methodology for model training and evaluation, with only the training data varying. Full implementation details for our data, model, and evaluation sets can be found in Appendix ~\ref{app:data} and ~\ref{app:exp}.

\paragraph{Data Generation.} 
To create controlled training sets for our experiments, we created two distinct pools of synthetic data:
\begin{itemize}[leftmargin=*, itemsep=0pt]
    \item For the source diversity experiments, we selected 8 distinct source datasets. For each source, we selected 10k real samples and applied 4 generators to synthesize 10k fake samples each, creating a pool of 40k fake samples per source.
    \item For generator diversity experiments, we selected 16 distinct generators. For each generator, we selected a pool of 40k fake samples derived from two fixed source datasets. 
\end{itemize}

\paragraph{Model Training.} 
To ensure our results are robust and not dependent on a specific random selection of sources or generators, we performed 3 independent runs for each experimental condition (e.g., for a specific value of $N_S$ and $V$). Each run used a different random combination of the available sources or generators.

\paragraph{Evaluation Protocol.}
We measure generalization on 10 out-of-domain test sets. For each model, we compute the macro-average Equal Error Rate (EER) and Accuracy (ACC) across all 10 sets. Since ACC relies on a fixed threshold (0.5) while EER finds an optimal one, the two metrics capture different aspects of model performance. To provide a single, robust metric that balances both, we introduce the Calibrated Detection Error (CDE), defined as the harmonic mean of EER and (1-ACC):
\begin{equation}
    CDE=\frac{2\cdot EER\cdot(1-ACC)}{EER+(1-ACC)}
\end{equation}

\begin{figure*}[t]
    \centering
    \includegraphics[
    width=\textwidth,
    trim={1mm, 3mm, 3mm, 0mm}
    ]{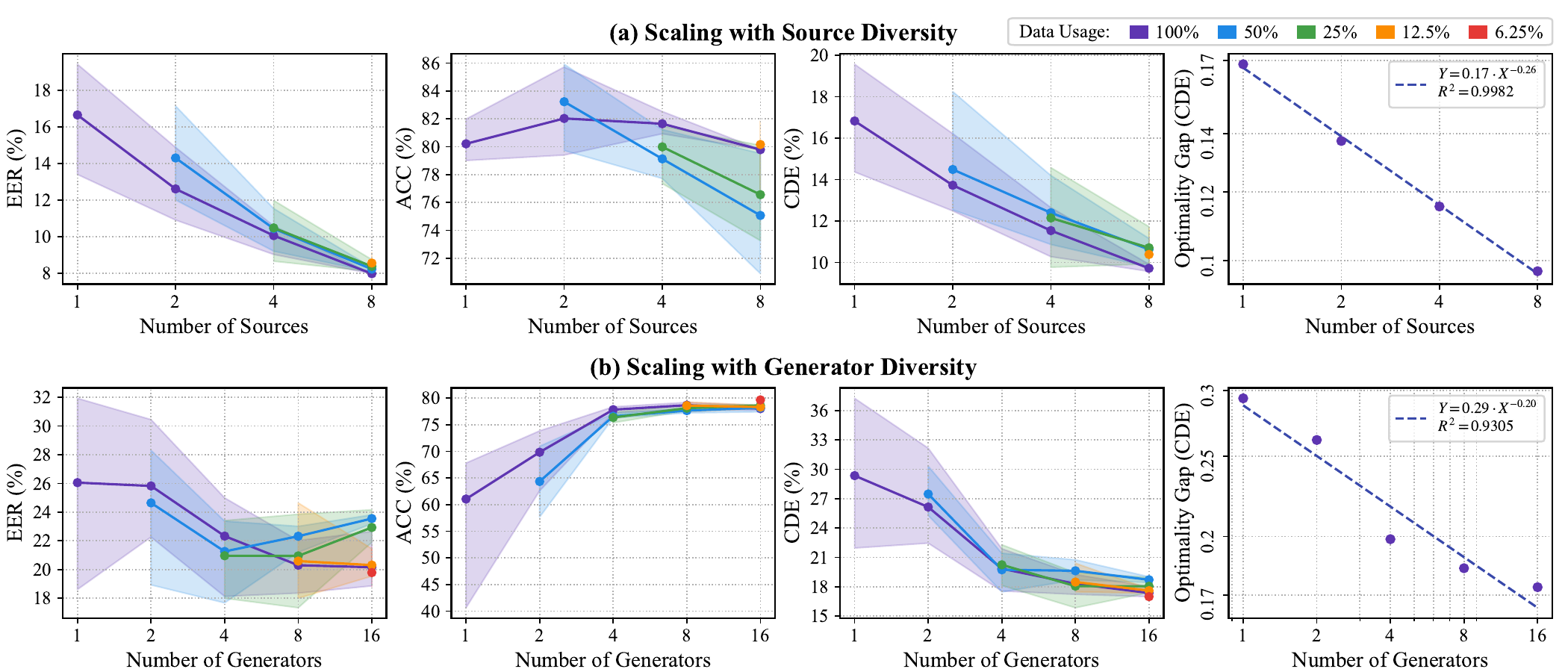}
    \caption{\textbf{Scaling with source and generator diversity.} Columns 1-3 plot generalization performance (EER, ACC, CDE) versus the number of training sources (a) or generators (b) on a logarithmic scale. Colored lines and shaded regions represent the average and min/max range, respectively, across experimental trials for different data usage percentages. Column 4 confirms a power-law fit for the Optimality Gap (CDE) at full data usage, with the fit equation and correlation coefficient provided in the legend.}
    \label{fig:scale}
\end{figure*}

\subsection{Empirical Analysis}
\label{sec:finding}
Figure~\ref{fig:scale} summarizes the results of our experiments on scaling with source and generator diversity. Our analysis of these results reveals three key findings that characterize the relationship between data composition and model generalization.

\paragraph{Finding 1:}
\textbf{Diversity is the primary driver of generalization.}
Our first key finding is that for a fixed data budget, increasing diversity both in sources and generators yields substantially greater performance gains than increasing the data volume from fewer diverse units. A model trained on 8 sources with 12.5\% data usage consistently outperforms a model trained on 1 source with 100\% usage. A similar trend is observed for generator diversity, where a model trained on 16 generators with just 6.25\% of the data  from each (2,500 samples per generator) achieves the best overall performance. This demonstrates the principle of diminishing returns. While increasing data volume is beneficial when diversity is low, its impact plateaus quickly beyond a saturation point. For data-efficient training, acquiring new information (increasing diversity) is a more effective strategy than reinforcing old information (increasing volume).


\paragraph{Finding 2:}
\textbf{The divergent roles of source and generator diversity.}
While both forms of diversity improve overall performance (CDE), a closer analysis of EER and ACC reveals that they play surprisingly different and complementary roles. 
\begin{itemize}[leftmargin=*, itemsep=0pt]
    \item \textbf{Source diversity improves discrimination.} Increasing the number of sources consistently lowers EER but can degrade ACC. We hypothesize this is because a variety of sources helps the model build a more robust manifold of genuine speech, enhancing the fundamental separability of the real and fake score distributions.
    \item \textbf{Generator diversity improves identification.} Conversely, increasing the number of generators strongly improves ACC, while its effect on EER is less consistent. We hypothesize this is because exposure to diverse forgery artifacts makes the model a better identifier, learning to confidently classify fakes but potentially at the cost of a less distinct real/fake decision boundary.
\end{itemize} 

\paragraph{Finding 3:}
\textbf{Generalization follows a predictable power Law.}
Finally, we find that despite the complex behaviors of individual metrics, overall generalization performance follows a predictable pattern. As shown in Figure~\ref{fig:scale}, the optimality gap (CDE) scales as a power law with both source and generator diversity, confirmed by high correlation coefficients (Pearson's $R^2$). This result demonstrates that generalization in SDD is not a random process but a principled, modelable phenomenon. We conclude that investing in data diversity represents an efficient and effective strategy for improving detector robustness.

\section{Principled Data Mixing with DOSS}
\label{sec:doss}

Our scaling law analysis provides a clear prescription: maximize data diversity. In practice, the most straightforward path to achieving this is to aggregate the numerous datasets. This action, however, creates a heterogeneous data pool with significant imbalances, directly raising the challenge posed in RQ2: what is the most effective and efficient strategy for mixing this data to best utilize its diversity?

Inspired by the principle of maximum entropy, we argue that for robust generalization against unpredicted future attacks, there is no prior evidence that any single known attack is more important than another. The most robust and least-biased assumption, therefore, is that all attacks should be treated as equally important. This leads us to propose the \textit{Diversity-Optimized Sampling Strategy (DOSS)}, a framework designed to apply this principle to the realistic challenges of heterogeneous data mixtures.

\subsection{DOSS in Practice}
The DOSS framework operates on granular domains. We first index our entire heterogeneous data pool and define these domains as follows:
\begin{itemize}[leftmargin=*, itemsep=0pt]
    \item For real data, a domain is its source. We denote the set of all real domains as $\mathcal{R}$. 
    \item For fake data, a domain is the combination of its source and generator. We denote the set of all fake domains as $\mathcal{F}$. We use a function $\text{base}(f)$ to obtain the original source and to map the domains in $\mathcal{F}$ with the domains in $\mathcal{R}$.
\end{itemize}

DOSS approximates the uniform distribution through two strategies: a pruning method (DOSS-Select) and a re-weighting method (DOSS-Weight).

\begin{algorithm}[tb]
\caption{DOSS-Select}\label{alg:doss_s}
\begin{tabular}{l@{\hspace{2pt}}l}
\multicolumn{2}{l}{\textbf{Require:} Sets $\mathcal{F}, \mathcal{R}$; Size $n_d$ for each domain $d$;} \\ 
\multicolumn{2}{l}{\qquad\qquad Parameters $N_c, \rho$; Function $\text{base}(f)$} \\
\multicolumn{2}{l}{\textbf{Initialize:} $s$ as an empty map} \\
& // \textit{\textbf{Step 1:} Compute counts for fake domains} \\
\small{1:} & \textbf{for} $f \in \mathcal{F}$ \textbf{do} \\
\small{2:} & \quad $s[f] \gets \min(n_f, N_c)$ \\
\small{3:} & \textbf{end for} \\
& // \textit{\textbf{Step 2:} Compute counts for real domains} \\
\small{4:} & \textbf{for} $r \in \mathcal{R}$ \textbf{do} \\
\small{5:} & \quad $\Sigma_r \gets \sum_{f \in \mathcal{F} : \text{base}(f)=r} s[f]$ \\
\small{6:} & \quad $s[r] \gets \min(n_r, \Sigma_r \times \rho)$ \\
\small{7:} & \textbf{end for} \\
\small{8:} & \textbf{return} $s$ \quad \textit{// Return final domain counts} \\
\end{tabular}
\end{algorithm}

\begin{algorithm}[tb]
\caption{DOSS-Weight}\label{alg:doss_w}
\begin{tabular}{l@{\hspace{2pt}}l}
\multicolumn{2}{l}{\textbf{Require:} Sets $\mathcal{F}, \mathcal{R}$; Size $n_d$ for each domain $d$;} \\
\multicolumn{2}{l}{\qquad\qquad Parameters $N_c, \tau, \rho$; Function $\text{base}(f)$} \\
\multicolumn{2}{l}{\textbf{Initialize: } $s, w$ as empty maps} \\
& // \textit{\textbf{Step 1.} Compute weights for fake domains} \\
\small{1:} & \textbf{for} $f \in \mathcal{F}$ \textbf{do} \\
\small{2:} & \quad $s[f] \gets \min(n_f, N_c)$ \\
\small{3:} & \quad $w[f] \gets (s[f])^{1/\tau}$ \\
\small{4:} & \textbf{end for} \\
& // \textit{\textbf{Step 2.} Compute weights for real domains} \\
\small{5:} & \textbf{for} $r \in \mathcal{R}$ \textbf{do} \\
\small{6:} & \quad $\Sigma_r \gets \sum_{f \in \mathcal{F} : \text{base}(f)=r} s[f]$ \\
\small{7:} & \quad $s[r] \gets \min(n_r, \Sigma_r \times \rho)$ \\
\small{8:} & \quad $w[r] \gets (s[r])^{1/\tau}$ \\
\small{9:} & \textbf{end for} \\
& // \textit{\textbf{Step 3.} Enforce global real-to-fake ratio} \\
\small{10:} & $W_{\mathcal{F}} \gets \sum_{f \in \mathcal{F}} w[f]$;  $W_{\mathcal{R}} \gets \sum_{r \in \mathcal{R}} w[r]$ \\
\small{11:} & $\alpha_{\text{adj}} \gets (W_{\mathcal{F}} \times \rho) / W_{\mathcal{R}}$ \\
\small{12:} & \textbf{for} $r \in \mathcal{R}$ \textbf{do} \\
\small{13:} & \quad $w[r] \gets w[r] \times \alpha_{\text{adj}}$ \\
\small{14:} & \textbf{end for} \\
\small{15:} & \textbf{return} $w$ \quad \textit{// Return final domain weights} \\
\end{tabular}
\end{algorithm}

\paragraph{DOSS-Select.}
DOSS-Select is a data pruning strategy that creates a smaller, more balanced, and data-efficient training subset. This method is directly motivated by our finding in Section~\ref{sec:scale} that data volume provides diminishing returns beyond a saturation point. We denote this saturation cap as $N_c$. As detailed in Algorithm~\ref{alg:doss_s}, the process first determines the number of samples to select from each fake domain by capping its original size at $N_c$. Then, for each real domain, it selects a proportional number of real samples based on the aggregated fake count and ratio $\rho$, ensuring that the volume of real speech scales dynamically to preserve a consistent local class balance. The output is a list of counts for each domain, which guides the construction of the final pruned dataset.

\paragraph{DOSS-Weight.}
DOSS-Weight is a re-weighting strategy that uses the entire data pool but adjusts the sampling probability of each domain at training time. This allows the model to see data from smaller domains more frequently without discarding any samples. The process, detailed in Algorithm~\ref{alg:doss_w}, involves three steps. First, it calculates an initial weight for each fake domain by capping its size at $N_c$ and applying a diversity temperature $\tau$. Second, it calculates weights for real domains using the same proportional logic as DOSS-Select. Finally, it computes a global adjustment factor to scale all real domain weights, ensuring that the total sampling probability across all real and fake domains strictly adheres to the target ratio $\rho$. The output is a list of final domain weights used to guide a weighted random sampler during training.

\begin{table*}[ht]
\centering
\small 
\setlength{\tabcolsep}{4.5pt} 
\caption{\textbf{Generalization performance in EER\% ($\downarrow$) for different data selection and mixing strategies.} Within each strategy section, the best and second-best results per column are in \textbf{bold} and \underline{underline}. The overall top two results across all experiments are highlighted with \setlength{\fboxsep}{1.5pt}\colorbox{grey300}{darker} and \setlength{\fboxsep}{1.5pt}\colorbox{grey100}{lighter} grey backgrounds.}
\label{tab:main_eer}
\begin{tabularx}{\textwidth}{@{}l|c|*{1}{Y}|*{10}{Y}@{}}
\toprule
\textbf{Strategy} & \textbf{\#Hours} & \textbf{AVG} & \textbf{ASV19} & \textbf{DECR} & \textbf{ITW} & \textbf{SC} & \textbf{FOR} & \textbf{EF} & \textbf{ADD22} & \textbf{ADD23} & \textbf{CFAD} & \textbf{ODSS} \\
\midrule
\rowcolor{green50}\multicolumn{13}{@{}l}{\textit{\textbf{Traditional Benchmarks: }Training on single datasets}} \\
ADD22 & 24 & 28.23 & 32.73 & 37.98 & 20.63 & \underline{15.77} & 44.60 & 1.90 & 15.17$^\dagger$ & 45.68 & 25.19 & 42.61 \\
ASV19 & 25 & \textbf{12.06} & \textbf{0.26}$^{\dagger\ast}$   & \underline{9.19} & \textbf{7.50} & 20.24 & \underline{4.59} & \textbf{0.14} & 11.80 & \textbf{16.27} & \underline{21.01} & 29.63 \\
FOR & 48 & \underline{13.53} & 3.24 & 31.49 & 15.39 & \textbf{14.68} & \textbf{0.38}$^\dagger$ & 2.43 & \textbf{4.19} & 20.23 & \textbf{15.23} & \underline{28.07} \\
DECR & 58 & 19.25 & \underline{0.39} & \cellcolor{grey300}\textbf{0.02}$^{\dagger\ast}$ & \underline{7.85} & 47.93 & 65.06 & \underline{1.74} & \underline{9.58} & \underline{20.03} & 22.49 & \textbf{17.43} \\
\midrule
\rowcolor{blue50}\multicolumn{13}{@{}l}{\textit{\textbf{Baseline: }Naive aggregation with $k$ datasets}} \\
$k=2$	& 0.9k & 8.78 & \underline{0.16}$^\ast$ & 11.29 & 4.32 & 9.76 & 3.04 & \cellcolor{grey100}\underline{0.03} & 8.77 & 19.55 & 23.65 & 7.27 \\
$k=6$	& 1.3k & 6.45 & \textbf{0.10}$^\ast$ & \textbf{0.13}$^\ast$ & 1.71 & 14.32 & 1.35 & \cellcolor{grey300}\textbf{0.01} & 2.09 & \underline{12.57} & 20.73 & 11.49 \\
$k=8$	& 3.3k & \underline{4.37} & 0.37$^\ast$ & \underline{0.19}$^\ast$ & \underline{1.57} & \cellcolor{grey100}\underline{0.05} & \underline{0.93} & 0.57 & \underline{1.94} & 16.18 & \underline{15.72} & \textbf{6.20} \\
$k=12$& 6.4k & \textbf{3.29} & 0.19$^\ast$ & 0.31$^\ast$ & \cellcolor{grey100}\textbf{1.21} & \cellcolor{grey300}\textbf{0.05} & \textbf{0.72} & 0.29 & \textbf{1.94} & \textbf{6.44} & \textbf{15.14} & \underline{6.57} \\
\midrule
\rowcolor{indigo50}\multicolumn{13}{@{}l}{\textit{\textbf{DOSS-Select:} Pruning via saturation cap ($N_c$)}} \\
$N_c=100$ & 40 & 3.43 & 1.31$^\ast$ & 0.67$^\ast$ & 1.63 & 2.08 & 0.93 & \underline{0.11} & 1.54 & 6.70 & \underline{14.49} & 4.88 \\
$N_c=500$& 0.2k & 2.77 & 0.32$^\ast$ & \underline{0.35}$^\ast$ & 1.34 & 0.83 & 0.17 & \textbf{0.11} & 1.63 & \underline{4.42} & 14.69 & \cellcolor{grey100}\textbf{3.85} \\
$N_c=2500$& 0.8k & \textbf{2.69} & \textbf{0.18}$^\ast$ & \textbf{0.17}$^\ast$ & \underline{1.24} & \underline{0.47} & \cellcolor{grey300}\textbf{0.06} & {0.15} & \cellcolor{grey100}\textbf{1.26} & \textbf{4.22} & 15.27 & \underline{3.90} \\
$N_c=12500$ & 2.9k & \underline{2.72} & \underline{0.21}$^\ast$ & 0.36$^\ast$ & \textbf{1.23} & \textbf{0.21} & \underline{0.13} & {0.15} & \underline{1.40} & 4.90 & \textbf{14.20} & 4.39 \\
\midrule
\rowcolor{purple50}\multicolumn{13}{@{}l}{\textit{\textbf{DOSS-Weight:} Re-weighting via saturation ($N_c$) \& temperature ($\tau$)}} \\
$N_c=50000, \tau=1$ & 6.4k & 2.81 & 0.20$^\ast$ & 0.50$^\ast$ & {1.23} & \textbf{0.16} & {0.13} & 0.17 & 1.52 & 4.84 & 14.69 & 4.68 \\
$N_c=2500, \tau=1$ & 6.4k & 2.51 & {0.12}$^\ast$ & \underline{0.15}$^\ast$ & \underline{1.22} & 0.29 & 0.17 & \underline{0.09} & \underline{1.31} & 3.39 & \cellcolor{grey300}\textbf{13.00} & 5.35 \\
$N_c=2500, \tau=5$ & 6.4k & \cellcolor{grey300}\textbf{2.34} & \cellcolor{grey300}\textbf{0.09}$^\ast$ & \cellcolor{grey100}\textbf{0.13}$^\ast$ & \cellcolor{grey300}\textbf{1.16} & 0.24 & \cellcolor{grey100}\textbf{0.08} & 0.11 & \cellcolor{grey300}\textbf{1.25} & \cellcolor{grey100}\underline{2.97} & {13.91} & \cellcolor{grey300}\textbf{3.44} \\
$N_c=2500, \tau=100$ & 6.4k & \cellcolor{grey100}\underline{2.41} & \cellcolor{grey100}\underline{0.10}$^\ast$ & 0.16$^\ast$ & {1.23} & \underline{0.21} & \underline{0.08} & \textbf{0.07} & 1.40 & \cellcolor{grey300}\textbf{2.56} & \cellcolor{grey100}\underline{13.89} & \underline{4.42} \\
\bottomrule
\end{tabularx}
\centering
\vspace{1mm}
\begin{tabular}{@{}l@{}}
\footnotesize
$\ast$: in-domain test set; $\dagger$: test set corresponding to the original traditional benchmark; all other columns are out-of-domain.
\end{tabular}
\end{table*}

\subsection{Experimental Validation}

To validate the proposed DOSS framework, we conduct a comparative study using a fixed model architecture across different data selection and mixing strategies. Generalization performance is evaluated on 10 distinct test sets, with the primary results shown in Table~\ref{tab:main_eer}. Full details of the experimental setup can be found in Appendix~\ref{app:data} and ~\ref{app:exp}.

\paragraph{Traditional Benchmarks.}
We first examine the performance of models trained under the traditional paradigm: using a single dataset for training and evaluating on both in-domain and out-of-domain test sets. The results clearly show that models trained this way are brittle and fail to generalize effectively. While they often achieve excellent performance on their corresponding in-domain test sets, their performance degrades significantly on out-of-domain data. For instance, the model trained only on DECR achieves a near-perfect 0.02\% EER on its own test set but has a poor average EER of 19.25\% across all ten sets. This demonstrates that single-dataset training encourages the model to overfit to dataset-specific biases rather than learning a universal representation of artificiality.

\paragraph{Baseline.} 
Next, we establish a baseline by naively aggregating an increasing number of training datasets ($k$). Simply increasing the size and diversity of the training pool leads to a substantial improvement in overall generalization, with the average EER dropping from 8.78\% for $k$=2 to 3.29\% for $k$=12. However, this `brute force' approach has a notable flaw. In such a heterogeneous and imbalanced pool, the training process is naturally dominated by the larger datasets, causing the model to prioritize the most common features. This can lead to negative transfer, where the model de-prioritizes artifacts from smaller domains, degrading performance on specific test sets as more data is added. For instance, performance on the ADD23 test set worsens from an EER of 12.57\% when using $k$=6 datasets to 16.18\% with $k$=8 datasets. This instability demonstrates that while naive aggregation is an improvement over single-dataset training, it is an unpredictable and suboptimal strategy, highlighting the need for a principled approach to data mixing.

\paragraph{DOSS-Select.} 
Our evaluation of DOSS-Select demonstrates that principled data pruning is highly efficient and outperforms naive aggregation. With a saturation cap of $N_c=100$ (40 hours of data), the model achieves a 3.43\% EER, which is better than all traditional benchmarks and most naive aggregation settings. Furthermore, increasing the cap to $N_c=500$ uses just 0.2k hours ($\approx$3\% of the data) but yields a 2.77\% EER. This result is particularly striking as it surpasses the best naive aggregation baseline that required the full 6.4k hours. This confirms that effective generalization is driven less by raw data volume and more by diverse, well-balanced composition.

The choice of $N_c$ is critical. Performance improves as the cap increases, reaching the overall best EER of 2.69\% at $N_c=2500$ (0.8k hours). However, increasing the cap further to $N_c=12500$ provides no additional benefit, with performance plateauing. We attribute this to two factors. First, the model has likely reached the data saturation point for learning from the available domains. Second, Figure~\ref{fig:domain}(a) in the Appendix shows that the domain distribution for $N_c=12500$ is less uniform than for $N_c=2500$, as the high cap re-introduces the natural imbalance of the original data pool. 


\paragraph{DOSS-Weight.} In contrast to pruning, our DOSS-Weight strategy re-weights the entire data pool and achieves the most effective results overall. The optimal setting ($N_c=2500, \tau=5$) yields an average EER of 2.34\%, the lowest across all tested methods. This corresponds to a relative error reduction of approximately 29\% compared to the best naive aggregation baseline and 13\% compared to the best DOSS-Select result. 

The choice of saturation cap $N_c$ remains critical. A poorly chosen $N_c$ can undermine the strategy's effectiveness. For instance, a large cap ($N_c=50000$) with no temperature balancing ($\tau=1$) creates a distribution similar to naive pooling (Figure~\ref{fig:domain}(b)) and performs worse than DOSS-Select. In contrast, a more reasonable cap like $N_c=2500$ creates a more uniform base distribution for the temperature to act upon.

The temperature parameter $\tau$ is key to refining this distribution. At $N_c=2500$, increasing $\tau$ from 1 to 5 improves performance from 2.51\% to 2.34\% EER. The temperature primarily balances the sampling probability among the real domains, which have more varied initial weights. However, increasing $\tau$ further to 100 provides no additional benefit, as the change in distribution is minimal. 

Ultimately, the performance advantage of DOSS-Weight over DOSS-Select suggests that re-weighting is a more powerful strategy than pruning. Instead of discarding potentially redundant data, it is better to retain it as intra-domain variations and down-weight its influence during training.

\begin{table*}[htb]
\centering
\small 
\setlength{\tabcolsep}{5pt} 
\caption{\textbf{Out-of-domain EER\% ($\downarrow$) comparison with prior works.} The State-of-the-Art (SOTA) results are collected from different systems in the literature. The best and second-best results are in \textbf{bold} and \underline{underline}.}
\label{tab:compare_eer}
\begin{tabularx}{\textwidth}{@{}l|c|c|Y|*{8}{Y}@{}}
\toprule
\textbf{System} & \textbf{\#Params} & \textbf{\#Hours} & \textbf{AVG} & \textbf{ITW} & \textbf{FOR} & \textbf{EF} & \textbf{ADD22} & \textbf{ADD23} & \textbf{ODSS} & \textbf{DV} & \textbf{FSW} \\
\midrule
\rowcolor{red50}\multicolumn{12}{@{}l}{\textit{\textbf{Existing Benchmarks} (Collected in Table~\ref{tab:sota})}} \\
SOTA & -- & -- & -- & 1.23 & 0.97 & \underline{0.20} & \underline{1.05} & 4.67 & \textbf{1.13} & 2.27 & 11.58 \\ \midrule
\rowcolor{blue50}\multicolumn{12}{@{}l}{\textit{\textbf{Training on Naive Aggregation} (\citealp{antideepfake_2025})}} \\
MMS-300M & 317M & 74k & 6.61 & 2.90 & 6.08 & 1.58 & 2.64 & 7.96 & 13.34 & 2.27 & 16.15 \\
MMS-1B & 965M & 74k & 7.24 & 1.82 & 1.73 & 0.34 & 2.76 & 9.05 & 5.49 & 2.47 & 23.81 \\
XLS-R-1B & 965M & 74k & 4.47 & 1.37 & 12.15 & 0.24 & 1.68 & 5.39 & 1.53 & 2.35 & 21.45 \\
XLS-R-2B & 2.2B & 74k & 3.94 & 1.23 & 1.73 & \underline{0.20} & \underline{1.05} & 4.67 & \textbf{1.13} & 2.35 & 19.14 \\
\midrule
\rowcolor{purple50}\multicolumn{12}{@{}l}{\textit{\textbf{Training with DOSS-Weight} (Ours)}} \\
XLS-R-300M & 317M & 12k & \underline{2.14} & \underline{0.81} & \underline{0.25} & 0.29 & \textbf{0.82} & \underline{3.63} & 1.65 & \underline{0.97} & \underline{8.70} \\
XLS-R-1B & 965M & 12k & \textbf{1.65} & \textbf{0.80} & \textbf{0.13} & \textbf{0.10} & 1.40 & \textbf{2.25} & \underline{1.23} & \textbf{0.86} & \textbf{6.47} \\
\bottomrule
\end{tabularx}
\end{table*}

\section{Scalability and Application of DOSS}
\label{sec:final}
This section validates the robustness and scalability of the DOSS framework. We expand our data pool, apply DOSS to train final models, and evaluate them against state-of-the-art methods.

\subsection{Data Pool Curation}
To build a comprehensive training set and test our framework's scalability, we expanded our data pool using the following data curation pipeline:

\begin{enumerate}[leftmargin=*, itemsep=0pt]
    \item \textbf{Collection}: We aggregated 17 publicly available datasets, unified their audio formats, and gathered metadata on their respective sources and generators.
    \item \textbf{Reorganization}: We de-duplicated the pool by identifying fake datasets that shared common real source corpora (e.g., VCTK, LibriTTS) and replaced the redundant real audio with canonical source versions.
    \item \textbf{Enrichment}: We analyzed the pool for gaps in generator diversity and synthesized new data from recent models to ensure our final pool reflects the current state of speech synthesis.
\end{enumerate}

This curation process resulted in a new 12k-hour data pool with enhanced diversity and reduced redundancy. Full details are in Appendix~\ref{app:train_sec5}.

\subsection{Final Model Performance}

Using the new data pool, we train final models (based on XLS-R~\cite{babu2021xlsr}, details in~\ref{app:exp}) using the DOSS-Weight strategy.
To validate its performance, we conducted a comprehensive evaluation against both established public benchmarks and a new set of commercial APIs.

\paragraph{Performance on Public Benchmarks.}
Table~\ref{tab:compare_eer} presents the out-of-domain evaluation on 8 public test sets. Our results highlight two findings regarding data efficiency and model scaling.

First, we compare our data-centric approach against the large-scale baselines established by~\citealp{antideepfake_2025}. Their best-performing system utilizes a massive XLS-R-2B backbone trained on 74k hours of data to achieve an average EER of 3.94\%. In contrast, our DOSS-trained model, using a smaller XLS-R-300M model and our 12k-hour curated data pool, achieves a significantly lower average EER of 2.14\%. This result highlights a clear efficiency advantage. By prioritizing generator diversity over raw data volume, we surpass the performance of a model with roughly 7 times more parameters trained on 6 times more data.

Second, we observe robust scaling behavior within our framework. Scaling the backbone from 300M to 1B parameters yields further performance gains and reduces the average EER by approximately 23\%. Notably, our 1B model establishes a new state-of-the-art on 6 out of the 8 individual benchmarks, surpassing the aggregated best results from prior literature. This suggests that the increased capacity allows the model to better capture the diverse artifacts present in the training data.

\begin{table*}[htbp]
\centering
\small
\setlength{\tabcolsep}{3pt}
\caption{\textbf{Detection ACC\% ($\uparrow$) on commercial APIs.} Results are evaluated on 5,000 synthetic samples per API (2,500 English + 2,500 Chinese). The best and second-best results are in \textbf{bold} and \underline{underline}.}
\label{tab:api}
\begin{tabularx}{\linewidth}{@{}l|Y|*{9}{Y}@{}}
\toprule
\textbf{System} & \textbf{AVG} & \textbf{Google} & \textbf{Microsoft} & \textbf{OpenAI} & \textbf{11Labs} & \textbf{Alibaba} & \textbf{Baidu} & \textbf{iFlytek} & \textbf{MiniMax} & \textbf{Qwen3} \\
\midrule
\rowcolor{blue50}\multicolumn{11}{@{}l}{\textit{\textbf{Training on Naive Aggregation} (\citealp{antideepfake_2025})}} \\
XLS-R-2B & 86.31 & 99.54 & 85.42 & \textbf{99.20} & 81.90 & 96.06 & 98.96 & 99.34 & 76.48 & 39.88\\
\midrule
\rowcolor{purple50}\multicolumn{11}{@{}l}{\textit{\textbf{Training with DOSS-Weight} (Ours)}} \\
XLS-R-300M & \underline{92.81} & \underline{99.96} & \underline{99.84} & 98.04 & \textbf{92.12} & \underline{98.92} & \textbf{100.00} & \textbf{99.98} & \underline{90.50} & \underline{55.94} \\
XLS-R-1B & \textbf{96.01} & \textbf{99.98} & \textbf{99.92} & \underline{98.92} & \underline{86.30} & \textbf{99.78} & \underline{99.98} & \underline{99.96} & \textbf{91.94} & \textbf{87.32} \\
\bottomrule
\end{tabularx}
\end{table*}

\paragraph{Performance on Commercial APIs.}
To further evaluate our model's performance against current, real-world threats, we created a new challenge set using 9 different commercial APIs. This complements our previous results on public, open-source test sets, which may not represent the latest generation of synthesis technology. We generated 2,500 synthetic samples from the latest version of each API in both English and Chinese (full details in Appendix~\ref{app:api}) and evaluated our model alongside the large-scale baseline from~\citealp{antideepfake_2025}. 

Table~\ref{tab:api} reports the overall detection accuracy. Our DOSS-trained models exhibit remarkable generalization compared to the naive baselines. While the massive XLS-R-2B baseline achieves an average accuracy of 86.31\%, our XLS-R-1B model improves this by nearly 10 absolute percentage points to 96.01\%.
This advantage is most visible on the most challenging, high-fidelity synthesis systems. For instance, on the Qwen3 TTS where the baseline accuracy collapses to 39.88\%, our 1B model maintains a high accuracy of 87.32\%. Similarly, on MiniMax, our model improves detection from 76.48\% to 91.94\%. This suggests that the wild diversity in our training set transfers effectively to unseen, advanced commercial systems. Note that performance varies by language for certain providers; a detailed breakdown of these linguistic differences is provided in Table~\ref{tab:api_full} in Appendix~\ref{app:other}.

\vspace{-2mm}
\section{Conclusion}
\vspace{-2mm}
This paper proposes a data-centric approach to generalization in speech deepfake detection. We first conducted a large-scale empirical study demonstrating that source and generator diversity are more impactful for generalization than raw data volume. 
Guided by these scaling laws, we introduced the Diversity-Optimized Sampling Strategy (DOSS) to effectively manage heterogeneous data mixtures. Our experiments validate that this principled approach enables both high data efficiency through pruning and superior representation through re-weighting, proving that smart data composition is superior to naive aggregation. Consequently, our final model establishes a new state-of-the-art, surpassing large-scale baselines on both public benchmarks and a commercial API challenge set with significantly greater data and model efficiency.


\section*{Limitations}
Our work focuses on generalization from a data-centric perspective. To isolate the effects of data composition, we primarily utilized a fixed model architecture and training configuration. Consequently, while we validated the scalability of our approach by extending from 300M to 1B parameters, this study does not explore the exhaustive interplay with other factors like massive model scaling or varied computational budgets. Investigating how these data scaling laws and mixing strategies co-adapt across a broader spectrum of model sizes remains a valuable direction for future work.

Furthermore, our curated data pool is linguistically concentrated on English and Chinese. While this reflects the composition of most publicly available datasets, it could limit our model's generalization to other languages. Building a multi-lingual detector is a significant future challenge that would require not only the collection of rare, diverse-language datasets but also principled methods to manage the resulting linguistic imbalance.


\section*{Acknowledgements}
The authors would like to thank Prof. Shinji Watanabe from CMU for his insightful discussions on this work.


\clearpage
\appendix

\section{Dataset Details}
\label{app:data}

\subsection{Training Data for Section~\ref{sec:scale}}
\label{app:text}
\paragraph{Source Diversity.}
To investigate the impact of source diversity, we first selected 8 distinct source speech datasets, comprising 4 English (EN) and 4 Chinese (ZH): 
\begin{itemize}[leftmargin=*, itemsep=0pt]
    \item \textbf{EN:} VCTK~\cite{veaux2013voice}, LibriTTS~\cite{zen2019libritts}, MLS~\cite{pratap2020mls}, and CommonVoice~\cite{ardila2019common}.
    \item \textbf{ZH:} Aishell1~\cite{bu2017aishell}, Aishell3~\cite{shi2020aishell}, MagicData~\cite{yang2022open}, and CommonVoice~\cite{ardila2019common}.
\end{itemize}

From each of these 8 datasets, we selected 10k audio samples to serve as the bonafide (real) data and as the basis for the generation process. Next, to create the synthetic counterparts, we chose 4 zero-shot TTS models: MaskGCT~\cite{wang2024maskgct}, F5TTS~\cite{chen2024f5}, E2TTS~\cite{eskimez2024e2}, and CosyVoice2~\cite{du2024cosyvoice}. For each real source dataset, we randomly selected speech prompts and text transcripts from it, using each of the 4 TTS models to generate 10,000 fake samples. This process yielded a total of 80k real samples (8 sources $\times$ 10k) and 320k fake samples (8 sources $\times$ 4 generators $\times$ 10k).

To form the training sets, we maintained a fixed real-to-fake ratio ($\rho=0.25$). We defined two scaling variables: the number of source datasets ($N_S \in \{1, 2, 4, 8\}$) and a data usage percentage ($V \in \{100\%, 50\%, 25\%, 12.5\%\}$). For any given configuration ($N_S, V$), the training set was constructed by aggregating $10\text{k} \times V$ real samples and $4 \times 10\text{k} \times V$ fake samples (from the 4 generators) from each of the $N_S$ sources. By varying $N_S$ and $V$, we created 10 distinct experimental settings (as shown in Figure~\ref{fig:scale}(a)), resulting in a minimum training set size of 50k samples.

\paragraph{Generator Diversity.}
To investigate the impact of generator diversity, we leverage the data provided in SpeechFake~\cite{huang2025speechfake}. We first select 2 source datasets, VCTK~\cite{veaux2013voice} and LibriTTS~\cite{zen2019libritts}, and establish a real data pool of 80k samples from each. Next, we select a total of 16 generators from SpeechFake that were built from these two sources, including models from three categories: 7 TTS (text-to-speech with fixed voice), 3 VC (voice clone), and 6 NV (neural-vocoded speech):
\begin{itemize}[leftmargin=*, itemsep=0pt]
    \item \textbf{TTS:}  GlowTTS~\cite{kim2020glow}, ProDiff-TTS~\cite{huang2022prodiff}, DiffGAN-TTS~\cite{liu2022diffgan}, TorToiSe~\cite{betker2023better}, 
    MeloTTS\footnote{\url{https://github.com/myshell-ai/MeloTTS}}, ChatTTS\footnote{\url{https://github.com/2noise/ChatTTS}}, CosyVoice~\cite{du2024cosyvoice} 
    \item \textbf{VC: } CosyVoice~\cite{du2024cosyvoice}, OpenVoice~\cite{qin2023openvoice} and its variant.
    \item \textbf{NV:} MelGAN~\cite{kumar2019melgan}, ParallelWaveGAN~\cite{yamamoto2020parallel}, HifiGAN~\cite{kong2020hifi}, FullBandMelGAN~\cite{yang2021multi}, FastDiff~\cite{huang2022fastdiff}, BigVGAN~\cite{lee2023bigvgan}
\end{itemize}

From each of these 16 generators, we formed a data pool of 40k fake samples (20k sourced from VCTK and 20k from LibriTTS). To form the training sets, we maintained a fixed real-to-fake ratio ($\rho=0.25$). We defined two scaling variables: the number of source datasets ($N_G \in \{1, 2, 4, 8, 16\}$) and a data usage percentage ($V \in \{100\%, 50\%, 25\%, 12.5\%, 6.25\%\}$). For any given configuration ($N_G, V$), the training set was constructed by aggregating $40\text{k} \times V$ fake samples from each of the $N_G$ generators, along with a corresponding $10\text{k}\times V \times N_G$ real samples drawn from the real data pool. By varying $N_G$ and $V$, we created 15 distinct experimental settings (as shown in Figure~\ref{fig:scale}(b)), resulting in a minimum training set size of 50k samples.

\begin{table}[tb]
\centering
\small
\setlength{\tabcolsep}{12pt} 
\caption{\textbf{Overview of the publicly available datasets composing the training pool for the DOSS experiments} (Section~\ref{sec:doss}).}
\label{tab:doss_train}
\begin{tabularx}{\linewidth}{@{}llr@{}}
\toprule
\textbf{ID} & \textbf{Dataset} & \textbf{\#Hours} \\
\midrule
T01 & ASVspoof2019~\cite{wang2020asvspoof} & 48 \\
T02 & ASVspoof5~\cite{wang2024asvspoof} & 878 \\
T03 & ADD2022~\cite{yi2022add} & 54 \\
T04 & ADD2023~\cite{yi2024add} & 49 \\
T05 & DFADD~\cite{du2024dfadd} & 227 \\
T06 & DECRO~\cite{ba2023transferring} & 94 \\
T07 & DiffSSD~\cite{bhagtani2025diffssd} & 58 \\
T08 & SpoofCeleb~\cite{jung2025spoofceleb} & 1925 \\
T09 & SpeechFake~\cite{huang2025speechfake} & 2837 \\
T10 & EmoFake~\cite{zhao2024emofake} & 28 \\
T11 & FoR~\cite{reimao2019dataset} & 48 \\
T12 & CFAD~\cite{ma2024cfad} & 109 \\
\bottomrule
\end{tabularx}
\end{table}

\begin{table*}[htb]
\centering
\small
\setlength{\tabcolsep}{5pt}
\caption{\textbf{Overview of the datasets used in the data pool curation} (Section~\ref{sec:final}). \#Dom., \#Src., and \#Gen. denote the number of domains, sources, and generators, respectively. * represent unknown generator.}
\label{tab:pool}
\begin{tabularx}{\textwidth}{@{}l l l l r r c r@{}}
\toprule
\textbf{ID} & \textbf{Datasets} & \textbf{Association} & \textbf{Language} & \textbf{\#Dom.} & \textbf{\#Src.} & \textbf{\#Gen.} & \textbf{\#Hours} \\
\midrule
\rowcolor{red50}\multicolumn{4}{@{}l}{\textit{\textbf{Fake Datasets }}} & \textbf{332} & \textbf{18} & \textbf{175} & \textbf{9717} \\
F01 & ADD2022~\cite{yi2022add} & R02 & zh & 1 & 1 & 1* & 44 \\
F02 & ADD2023~\cite{yi2024add} & R02 & zh & 7 & 1 & 7 & 49 \\
F03 & ASVspoof2019~\cite{wang2020asvspoof} & R11 & en & 19 & 1 & 19 & 105 \\
F04 & ASVspoof2021~\cite{yamagishi2021asvspoof} & R11 & en & 17 & 1 & 17 & 620 \\
F05 & ASVspoof5~\cite{wang2024asvspoof} & R10 & en & 32 & 1 & 32 & 1825 \\
F06 & CDADD~\cite{li2024cross} & R07 & en & 5 & 1 & 5 & 268 \\
F07 & CFAD~\cite{ma2024cfad} & R02 & zh & 12 & 1 & 12 & 132 \\
F08 & CodecFake~\cite{xie2025codecfake} & R02,R11 & en,zh & 14 & 2 & 7 & 885 \\
F09 & DECRO~\cite{ba2023transferring} & R04 & en, zh & 26 & 2 & 16 & 99 \\
F10 & DFADD~\cite{du2024dfadd} & R11 & en & 5 & 1 & 5 & 180 \\
F11 & DiffSSD~\cite{bhagtani2025diffssd} & R07-08 & en & 7 & 2 & 7 & 58 \\
F12 & EmoFake~\cite{zhao2024emofake} & R05 & en & 1 & 1 & 1* & 17 \\
F13 & FoR~\cite{reimao2019dataset} & R08 & en & 1 & 1 & 1* & 29 \\
F14 & FakeSpeechWild~\cite{xie2025fake} & R06 & zh & 4 & 4 & 1* & 41 \\
F15 & TIMIT-TTS~\cite{salvi2023timit} & R08,R11 & en & 12 & 1 & 12 & 17 \\
F16 & SpoofCeleb~\cite{jung2025spoofceleb} & R12 & en & 23 & 1 & 23 & 1816 \\
F17 & SpeechFake~\cite{huang2025speechfake} & R01,R02,R07,R11 & en,zh,etc & 102 & 7 & 33 & 3035 \\
F18 & Self-Generated & R01-03,R07,R09-11 & en,zh & 48 & 8 & 7 & 590 \\
\midrule
\rowcolor{green50}\multicolumn{4}{@{}l}{\textit{\textbf{Real Datasets }}} & \textbf{18} & \textbf{18} & \textbf{-} & \textbf{2356} \\
R01 & Aishell1~\cite{bu2017aishell} & F17-18 & zh & 1 & 1 & - & 39 \\
R02 & Aishell3~\cite{shi2020aishell} & F01-02,F07,F17-18 & zh & 1 & 1 & - & 86 \\
R03 & CommonVoice~\cite{ardila2019common} & F17-18 & en,zh,etc & 3 & 3 & - & 1023 \\
R04 & DECRO~\cite{ba2023transferring} & F09 & en,zh & 2 & 2 & - & 39 \\
R05 & ESD~\cite{zhou2022emotional} & F12 & en & 1 & 1 & - & 11 \\
R06 & FSW~\cite{xie2025fake} & F14 & zh & 4 & 4 & - & 33 \\
R07 & LibriTTS~\cite{zen2019libritts} & F06,F11,F17-18 & en & 1 & 1 & - & 245 \\
R08 & LJSpeech~\cite{ljspeech17} & F11,F13,F15 & en & 1 & 1 & - & 24 \\
R09 & MagicData & F18 & zh & 1 & 1 & - & 256 \\
R10 & MLS~\cite{pratap2020mls} & F05 & en & 1 & 1 & - & 350 \\
R11 & VCTK~\cite{veaux2013voice} & F03-04,F10,F15,F17-18 & en & 1 & 1 & - & 83 \\
R12 & Voxceleb~\cite{nagrani2017voxceleb} & F16 & en & 1 & 1 & - & 167 \\
\bottomrule
\end{tabularx}
\end{table*}

\subsection{Training Data for Section~\ref{sec:doss}}

Section~\ref{sec:doss} includes four parts, all conducted on the publicly available datasets detailed in Table~\ref{tab:doss_train}. We take the training and development splits from these datasets to form the data pool. The setup for each part is as follows:

\begin{itemize}[leftmargin=*, itemsep=0pt] 
    \item \textbf{Traditional Benchmarks:} We trained separate models on the individual training splits of four datasets: ASVspoof2019 (ASV19), ADD2022 (ADD22), DECRO (DECR), and FoR (FOR).
    \item \textbf{Baseline:} This experiment (and the following two) utilizes the full 12-dataset pool. We cumulatively aggregated the first $k$ datasets (from T01 to T$k$) to form training sets of increasing size:
    \begin{itemize}[leftmargin=*, itemsep=0pt]
        \item $k=2$: T01–T02 (926 hours)
        \item $k=6$: T01–T06 (1,351 hours)
        \item $k=8$: T01–T08 (3,335 hours)
        \item $k=12$: T01–T12 (6,357 hours, the full pool)
    \end{itemize}

    \item \textbf{DOSS-Select:} This strategy uses the full training pool ($k=12$) but selects a subset of samples from each domain based on the algorithm, resulting in a smaller, pruned dataset.
    \item \textbf{DOSS-Weight:} This strategy also uses the full training pool ($k=12$), but all samples are kept and assigned different sampling probabilities.
\end{itemize}

\subsection{Training Data for Section~\ref{sec:final}}
\label{app:train_sec5}
The final data pool used was curated through a three-step process, with full details in Table~\ref{tab:pool}.

First, we collected 17 publicly available SDD datasets. We unified their audio formats and gathered metadata on their respective sources and generators. This initial analysis revealed that many datasets, while large in volume, were limited in diversity (e.g., fewer than 20 domains).

Second, we reorganized and de-duplicated the pool. We traced the associations between fake datasets and their real source corpora, consolidating the shared real audio into a set of 12 datasets.

Third, we enriched the pool. An analysis of the aggregated generators revealed a gap in recent synthesis methods. We filled this gap by synthesizing new data from 7 recent generators: MegaTTS3~\cite{jiang2025megatts}, CosyVoice2~\cite{du2024cosyvoice}, Chatterbox\footnote{\url{https://
github.com/resemble-ai/chatterbox}}, MaskGCT~\cite{wang2024maskgct}, F5-TTS~\cite{chen2024f5}, E2-TTS~\cite{eskimez2024e2}, and IndexTTS~\cite{deng2025indextts}. Guided by our scaling law findings, we generated this new data using 8 different sources to ensure high acoustic diversity.

This curation process resulted in our final 12k-hour data pool, comprising 18 real domains and 332 fake domains.

\begin{table}[t]
\centering
\small
\setlength{\tabcolsep}{1.5pt} 
\caption{\textbf{Overview of the publicly available evaluation datasets used in experiments.} The `Notes' column specifies the exact subset (e.g., \texttt{LA\_eval}) or version (e.g., \texttt{normed}) used for evaluation. }
\label{tab:eval}
\begin{tabularx}{\linewidth}{@{}lll@{}}
\toprule
\textbf{Abbr.} & \textbf{Dataset} & \textbf{Notes} \\
\midrule
ASV19 & ASVspoof2019~\cite{wang2020asvspoof} & \texttt{LA\_eval} \\
DECR & DECRO~\cite{ba2023transferring} & \texttt{eval} \\
ITW & InTheWild~\cite{muller22does} & - \\
SC & SpoofCeleb~\cite{jung2025spoofceleb} & \texttt{eval} \\
FOR & FoR~\cite{reimao2019dataset}& \texttt{normed\_test} \\
EF & EmoFake~\cite{zhao2024emofake} & \texttt{eval} \\
ADD22 & ADD2022~\cite{yi2022add} & \texttt{Track3\_test2} \\
ADD23 & ADD2023~\cite{yi2024add} & \texttt{Track1\_testR2} \\
CFAD & CFAD~\cite{ma2024cfad} & \texttt{test\_unseen} \\
ODSS & ODSS~\cite{yaroshchuk2023open} & - \\
DV & DeepVoice~\cite{bird2023real} & \texttt{segmented} \\
FSW & FakeSpeechWild~\cite{xie2025fake} & \texttt{eval} \\
\bottomrule
\end{tabularx}
\end{table}

\subsection{Evaluation Datasets}
\paragraph{Public Benchmarks.}
Our experimental results are primarily reported on the public benchmarks detailed in Table~\ref{tab:eval}, which also lists the abbreviations used in our results. These evaluation sets can be broadly categorized by their relationship to the training data:
\begin{itemize}[leftmargin=*, itemsep=0pt]
    \item \textbf{In-domain (partially seen):} ASV19, the most common benchmark for SDD, shares some similar synthesis algorithms with its own training set. Similarly, the DECR evaluation set contains the same generators present in its training set.
    \item \textbf{Out-of-domain (unseen):} The other test sets represent more challenging, unseen scenarios. They consist of either entirely unknown generators (e.g., ITW, ADD22) or generators that were not part of their respective training data (e.g., SC, FOR, EF).
\end{itemize}
Notably, while most of these datasets consist of English or Chinese audio, ODSS provides multilingual evaluation, as it includes English, German, and Spanish samples.

The experiments in Section~\ref{sec:scale} and Section~\ref{sec:doss} use the first 10 test sets for evaluation. For Section~\ref{sec:final}, we modify this benchmark by excluding in-domain sets and complementing it with two others (e.g., DV and FSW).

\begin{table*}[t]
\centering
\small
\setlength{\tabcolsep}{6.5pt}
\caption{\textbf{Summary of commercial text-to-speech APIs used for generating synthetic speech data.}}
\label{tab:api_summary}
\begin{tabularx}{\textwidth}{@{}lllr@{}}
\toprule
\textbf{API} & \textbf{Provider} & \textbf{URL} & \textbf{Voices} \\
\midrule
Google Cloud TTS & Google & \url{https://cloud.google.com/text-to-speech} & 274 \\
Azure Speech Service & Microsoft & \url{https://learn.microsoft.com/azure/ai-services/speech-service} & 116 \\
GPT-4o mini TTS & OpenAI & \url{https://platform.openai.com/docs/guides/text-to-speech} & 10\\
ElevenLabs TTS & 11Labs & \url{https://Elevenlabs.io/} & 10 \\
Aliyun TTS & Alibaba & \url{https://ai.aliyun.com/nls/tts} & 90 \\
Baidu TTS & Baidu & \url{https://ai.baidu.com/tech/speech/tts} & 5 \\
Xfyun TTS & iFlytek & \url{https://www.xfyun.cn/services/online_tts} & 5 \\
MiniMax TTS & MiniMax & \url{https://www.minimaxi.com} & 41 \\
Qwen3 TTS Flash & Qwen3 & \url{https://help.aliyun.com/zh/model-studio/qwen-tts} & 17 \\
\bottomrule
\end{tabularx}
\end{table*}

\paragraph{Commercial APIs.}
\label{app:api}
To complement the generator types found in public benchmarks, we curated a new challenge set using 9 distinct commercial text-to-speech (TTS) APIs, as detailed in Table~\ref{tab:api_summary}. These platforms range from established cloud providers (e.g., Google, Microsoft) to cutting-edge generative engines (e.g., OpenAI, ElevenLabs, Qwen), ensuring broad coverage of modern acoustic qualities and synthesis architectures. We generated a total of 5,000 synthetic samples per API, split evenly between English and Chinese contexts (2,500 samples each). To maximize diversity, we utilized the full range of available voices on each platform. Furthermore, for APIs that support parameter customization, we enhanced variability by randomly altering attributes such as pitch, timbre, volume, and speaking rate. The source text was drawn from established real datasets as in~\ref{app:text}.

\section{Experimental Details}
\label{app:exp}

\subsection{Model Architecture}
For all experiments, we employ the self-supervised XLS-R~\cite{babu2021xlsr} architecture as our model backbone. This cross-lingual model was pre-trained on approximately 436k hours of unlabeled, publicly available speech data spanning 128 languages. This extensive, multilingual pre-training provides a powerful and generalized foundation for modeling speech representations and has been shown to achieve state-of-the-art performance in SDD~\cite{tak2022automatic}. 

We adapt this backbone for detection by adding a temporal average pooling layer and an MLP classifier head, fine-tuning the entire network on our training data. To balance computational efficiency with performance, we utilize the 300M-parameter version for the extensive empirical studies on scaling laws and data mixing (Sections~\ref{sec:scale} and~\ref{sec:doss}). For the final large-scale validation and robustness analysis (Section~\ref{sec:final}), we evaluate both the 300M and 1B-parameter versions to demonstrate the scalability of our approach.

\subsection{Training Configuration}
For all training, input audio is first resampled to 16kHz and then processed into 4-second segments; utterances shorter than this length are repeatedly padded, while longer utterances are randomly chunked. We apply two data augmentation techniques: 1) Rawboost~\cite{tak2022rawboost}: Applied with a probability of 0.5, this method introduces convolutional and additive noise directly to the raw waveform. 2) Codec Augmentation: Applied with a probability of 0.3, this simulates real-world compression artifacts by converting the audio to various formats (e.g., FLAC, MP3, AAC, Opus), improving robustness to different audio inputs. The total effective batch size is set to 128. 

We use the AdamW optimizer with a weight decay of 1e-4. The learning rate (LR) schedule was scaled by data volume: for our largest experiments (>1k hours), the LR was held constant at 1e-6 for the first 50k steps, then decayed exponentially to 1e-7 until 200k steps. For medium-scale experiments (100-1k hours), training was run for 100k steps. For small-scale experiments (<100 hours, e.g., experiments in Sec~\ref{sec:scale}), we trained for a fixed 50k steps to maintain a consistent training budget. We employ a weighted cross-entropy loss for training. The model is trained with a weighted cross-entropy loss, where the class weights are set based on the real-to-fake ratio to balance the dataset.

\subsection{Inference and Evaluation}
During inference, test audio is processed using the same 4-second segmentation strategy as in training. We evaluate performance using two primary metrics: Equal Error Rate (EER) and Accuracy (ACC). The EER is computed as a threshold-independent metric from the distribution of the raw real class scores. The ACC is calculated by taking the argmax of the final output predictions, which is equivalent to applying a fixed decision threshold of 0.5 to the softmax-normalized real score.

For our comparison with models from~\citealp{antideepfake_2025}, we used their publicly released checkpoint. Notably, to ensure a fair comparison, we followed its original inference protocol, which processes the full-length audio input directly, rather than applying our 4-second chunking strategy to avoid performance degradation on their specific models.

\section{Extended Experimental Results}
\label{app:other}
\subsection{Domain Distribution under DOSS}

\begin{figure}[htb]
    \centering
    \includegraphics[width=\linewidth, trim={0mm, 2mm, 0mm, 0mm}]{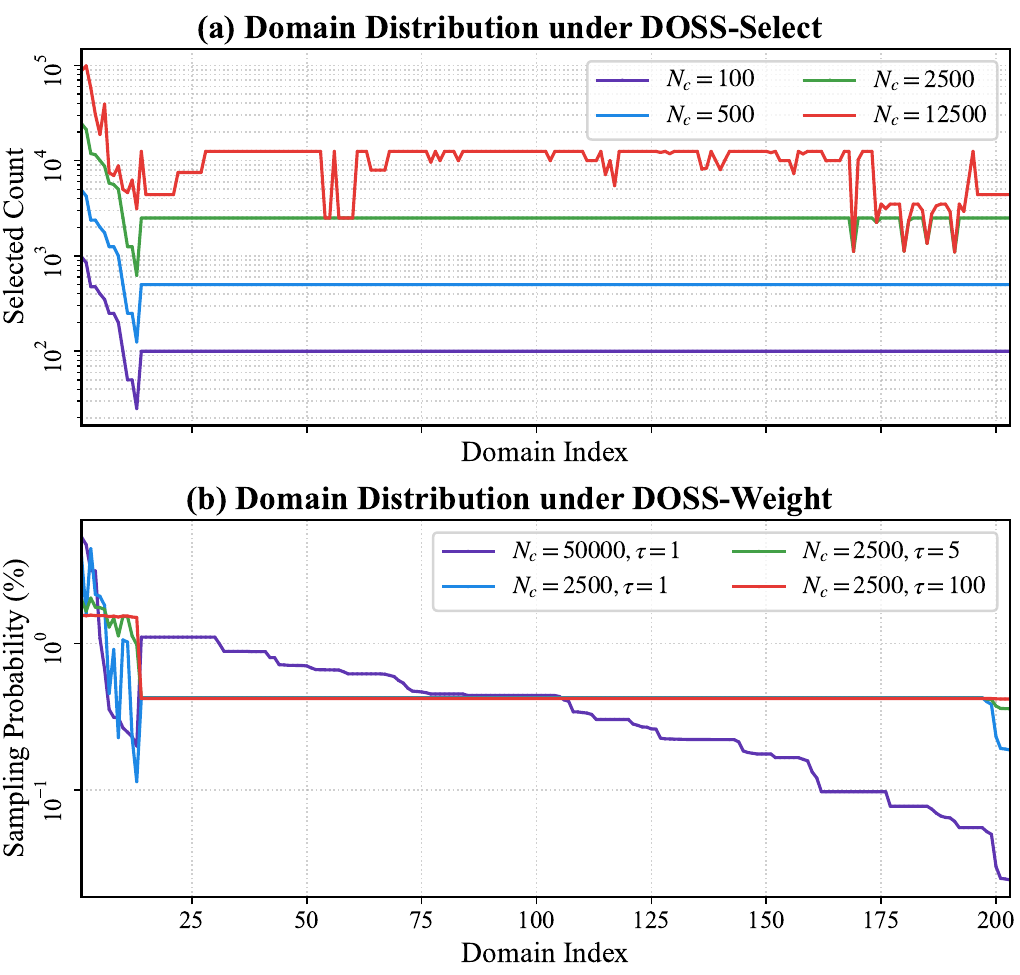}
    \caption{\textbf{Comparison of domain distributions under DOSS-Select and DOSS-Weight strategies.} Each colored line represents a distinct hyperparameter configuration as defined in Table~\ref{tab:main_eer}.}
    \label{fig:domain}
\end{figure}
Figure~\ref{fig:domain} illustrates the domain sampling distributions resulting from our two DOSS strategies. 
For DOSS-Select, the fake domain distribution is primarily determined by the saturation cap ($N_c$). By setting $N_c$ to a small value, we ensure that most fake domains are capped, which results in a near-uniform sampling distribution across them. However, since we use the proportional strategy, a real domain's sampling weight is derived from the sum of its associated fake domains. This leads to a non-uniform distribution for the real domains, as some are associated with many more fake domains than others.
For DOSS-Weight, the final distribution is controlled by both the saturation cap ($N_c$) and the temperature ($\tau$). If $N_c$ is set to a large value, the initial base importance of each domain remains highly varied. By setting a proper (smaller) $N_c$, we first create a near-uniform base importance for the fake domains, similar to DOSS-Select. Applying the temperature ($\tau$) then primarily serves to flatten the distribution of the real domains, allowing us to balance their weights.

\begin{table*}[htb]
\centering
\small 
\setlength{\tabcolsep}{5pt} 
\caption{\textbf{Out-of-domain ACC\% ($\uparrow$) comparison with prior works.} The best and second-best results are in \textbf{bold} and \underline{underline}.}
\label{tab:compare_acc}
\begin{tabularx}{\textwidth}{@{}l|c|c|Y|*{8}{Y}@{}}
\toprule
\textbf{System} & \textbf{\#Params} & \textbf{\#Hours} & \textbf{AVG} & \textbf{ITW} & \textbf{FOR} & \textbf{EF} & \textbf{ADD22} & \textbf{ADD23} & \textbf{ODSS} & \textbf{DV} & \textbf{FSW} \\
\midrule
\rowcolor{blue50}\multicolumn{12}{@{}l}{\textit{\textbf{Training on Naive Aggregation} (\citealp{antideepfake_2025})}} \\
MMS-300M & 317M & 74k & 89.60 & 97.31 & 87.21 & 98.39 & 89.58 & 94.08 & 86.59 & 90.49 & 73.13 \\
MMS-1B & 965M & 74k & 93.15 & 96.66 & \underline{98.05} & 98.22 & 97.55 & 93.11 & 95.01 & 96.44 & 70.21 \\
XLS-R-1B & 965M & 74k & 92.58 & 98.62 & 82.65 & 98.29 & 98.26 & 95.04 & \underline{98.46} & 92.43 & 76.92 \\
XLS-R-2B & 2.2B & 74k & 94.62 & \underline{98.70} & 96.48 & 98.45 & \underline{98.97} & 96.65 & \textbf{98.92} & 92.95 & 75.83 \\
\midrule
\rowcolor{purple50}\multicolumn{12}{@{}l}{\textit{\textbf{Training with DOSS-Weight} (Ours)}} \\
XLS-R-300M & 317M & 12k & \underline{96.88} & \textbf{98.82} & 97.84 & \underline{99.01} & \textbf{99.38} & \textbf{97.09} & 97.01 & \textbf{96.54} & \underline{89.31} \\
XLS-R-1B & 965M & 12k & \textbf{97.40} & 98.58 & \textbf{99.78} & \textbf{99.51} & 98.90 & \underline{96.93} & 96.34 & \underline{96.42} & \textbf{92.77} \\
\bottomrule
\end{tabularx}
\end{table*}

\begin{table}[t]
\centering
\small
\setlength{\tabcolsep}{6pt} 
\caption{\textbf{State-of-the-Art (SOTA) Benchmarks for Out-of-Domain Generalization.} This table lists the best-performing models from the literature for each test set, serving as the benchmark references for Table \ref{tab:compare_eer}.}
\label{tab:sota}
\begin{tabularx}{\linewidth}{@{}lllr@{}}
\toprule
\textbf{Test Set} & \textbf{SOTA System} & \textbf{Reference} & \textbf{EER(\%)} \\
\midrule
ITW & XLS-R-2B & \citealp{antideepfake_2025} & 1.23 \\
FOR & W2V-Large & \citealp{antideepfake_2025} & 0.97 \\
EF & XLS-R-2B & \citealp{antideepfake_2025} & 0.20 \\
ADD22 & XLS-R-2B & \citealp{antideepfake_2025} & 1.05 \\
ADD23 & XLS-R-2B & \citealp{antideepfake_2025} & 4.67 \\
ODSS & XLS-R-2B & \citealp{antideepfake_2025} & 1.13 \\
DV & MMS-300M & \citealp{antideepfake_2025} & 2.27 \\
FSW & XLSR-AASIST & \citealp{xie2025fake} & 11.58 \\
\bottomrule
\end{tabularx}
\end{table}

\begin{table*}[htbp]
\centering
\small
\setlength{\tabcolsep}{3pt}
\caption{\textbf{Detection ACC\% ($\uparrow$) on commercial APIs.} Results are evaluated on 2,500 synthetic samples per API. The table is vertically split by language test set (English vs. Chinese). The best and second-best results within each language section are in \textbf{bold} and \underline{underline}.}
\label{tab:api_full}
\begin{tabularx}{\linewidth}{@{}l|Y|*{9}{Y}@{}}
\toprule
\textbf{Model} & \textbf{AVG} & \textbf{Google} & \textbf{Microsoft} & \textbf{OpenAI} & \textbf{11Labs} & \textbf{Alibaba} & \textbf{Baidu} & \textbf{iFlytek} & \textbf{MiniMax} & \textbf{Qwen3} \\
\midrule
\multicolumn{11}{c}{\textbf{\textit{--- English (EN) Test Set ---}}} \\
\midrule
\rowcolor{blue50}\multicolumn{11}{@{}l}{\textit{Training on Naive Aggregation}} \\
MMS-300M & 76.36 & 98.40 & 83.92 & 91.40 & 45.00 & 90.36 & 99.40 & 98.52 & 69.00 & 11.28 \\
MMS-1B   & 79.40 & 99.60 & 90.36 & \textbf{98.72} & 38.76 & 97.48 & \underline{99.96} & 99.92 & 68.16 & 21.64 \\
XLS-R-1B & 82.79 & 99.16 & 91.00 & 97.20 & 65.08 & 93.00 & 99.88 & 98.44 & 68.96 & 32.36 \\
XLS-R-2B & 83.25 & 99.12 & 80.32 & \underline{98.52} & 63.97 & 96.44 & 99.76 & 99.64 & 74.36 & 37.12 \\
\midrule
\rowcolor{purple50}\multicolumn{11}{@{}l}{\textit{Training with DOSS-Weight (Ours)}} \\
XLS-R-300M & \underline{90.62} & \underline{99.92} & \underline{99.80} & 97.32 & \textbf{89.96} & \underline{98.00} & \textbf{100.00} & \textbf{99.96} & \textbf{89.32} & \underline{41.28} \\
XLS-R-1B   & \textbf{93.79} & \textbf{99.96} & \textbf{99.84} & 98.16 & \underline{82.60} & \textbf{99.56} & \textbf{100.00} & \underline{99.92} & \underline{87.68} & \textbf{76.40} \\
\midrule
\multicolumn{11}{c}{\textbf{\textit{--- Chinese (ZH) Test Set ---}}} \\
\midrule
\rowcolor{blue50}\multicolumn{11}{@{}l}{\textit{Training on Naive Aggregation}} \\
MMS-300M & 84.85 & 97.80 & 89.56 & 97.44 & 79.48 & 74.36 & 98.12 & 92.60 & 49.48 & 7.20 \\
MMS-1B   & 90.47 & 99.84 & 85.47 & 99.40 & 96.80 & 89.91 & 94.92 & \underline{99.64} & 57.80 & 22.28 \\
XLS-R-1B & 94.63 & 99.92 & 96.52 & 99.60 & \underline{99.80} & 89.56 & 98.92 & 99.36 & 73.40 & 53.84 \\
XLS-R-2B & 95.21 & \underline{99.96} & 90.52 & \textbf{99.88} & \textbf{99.84} & 95.68 & 98.16 & 99.04 & 78.60 & 42.64 \\
\midrule
\rowcolor{purple50}\multicolumn{11}{@{}l}{\textit{Training with DOSS-Weight (Ours)}} \\
XLS-R-300M & \underline{98.05} & \textbf{100.00} & \underline{99.88} & 98.76 & 94.28 & \underline{99.84} & \textbf{100.00} & \textbf{100.00} & \underline{91.68} & \underline{70.60} \\
XLS-R-1B   & \textbf{98.23} & \textbf{100.00} & \textbf{100.00} & \underline{99.68} & 90.00 & \textbf{100.00} & \underline{99.96} & \textbf{100.00} & \textbf{96.20} & \textbf{98.24} \\
\bottomrule
\end{tabularx}
\end{table*}

\subsection{Detailed Performance Analysis}
\paragraph{Convergence of Evaluation Metrics.}
Table~\ref{tab:compare_acc} provides the full Accuracy (ACC) results. Unlike the distinct scaling behaviors observed in Section~\ref{sec:scale} where EER and ACC occasionally diverged, the final results demonstrate that these metrics converge as overall model performance improves. This reflects a general phenomenon where, as the separation between real and fake distributions becomes robust, the performance disparity between an optimal threshold (EER) and a fixed threshold (ACC) naturally diminishes. Consistent with this trend, our final DOSS-trained systems achieve high average accuracy, with the 300M model reaching 96.88\% and the 1B model improving further to 97.40\%, mirroring the strong generalization observed in the EER analysis.

\paragraph{Fine-Grained Analysis of Commercial APIs.}
Table~\ref{tab:api_full} provides a granular breakdown of performance across 9 commercial providers, separated by language (English and Chinese). Analyzing the results reveals distinct tiers of difficulty among the APIs. Established providers such as Google, Microsoft, Baidu, and iFlytek appear to use synthesis methods that are readily detectable by all evaluated models, with accuracy scores consistently exceeding 98\% regardless of the training strategy. In contrast, the newer generation of generative TTS systems—specifically ElevenLabs, MiniMax, and Qwen3—poses a significantly greater challenge, causing performance drops across all models. Qwen3 proves to be the most difficult attacker in the English set, with the best-performing model (XLS-R-1B) achieving only 76.40\%, compared to $>$99\% on the easier APIs.

Furthermore, we observe a notable dependency on language, where detection performance for the same API can fluctuate drastically between English and Chinese test sets. For the baseline models trained on naive aggregation, this variance is extreme on ElevenLabs, where the XLS-R-2B model detects Chinese fakes with 99.84\% accuracy but fails on English fakes, dropping to 63.97\%. A similar language gap exists for our DOSS-trained models on the hardest benchmarks; for Qwen3, our XLS-R-1B model detects Chinese samples with 98.24\% accuracy but achieves only 76.40\% on the English samples. This confirms that synthesis differences exist between languages within these APIs, suggesting that variations in linguistic features or language-dependent vocoder artifacts can impact detector robustness.

\subsection{Latent Space Analysis} 
To investigate the internal representations learned by our final model (using XLS-R-300M as an example), we conduct an analysis of its latent space. We hypothesize that despite being trained on a simple binary (Real/Fake) objective, the model implicitly learns and encodes separable representations of intrinsic attributes, such as data source and generator.

\paragraph{Methodology} We test this hypothesis on a subset of our training data composed of 8 distinct sources and 8 distinct generators. Our analysis is two-fold:
\begin{itemize}[leftmargin=*, itemsep=0pt]
    \item \textbf{Quantitative Probing:} We extract embeddings from the temporal pooling layer. To ensure a fair, balanced comparison, we create two 10k-sample testbeds: one real (for source probing) and one fake (for source and generator probing). On each testbed, we train a simple linear classifier (Logistic Regression) as a "probe" to predict the (1) source ID or (2) generator ID from the frozen embeddings. The probe is trained on 80\% of the data and evaluated on a held-out 20\% test set over 5 random seeds.
    \item \textbf{Qualitative Visualization:} We use t-SNE to visualize the embeddings in 2D, coloring the points by their source or generator labels.
\end{itemize}

\paragraph{Results and Analysis}
\begin{figure}[htbp]
    \centering
    \includegraphics[width=\linewidth, trim={1mm, 5mm, 0mm, 0mm}]{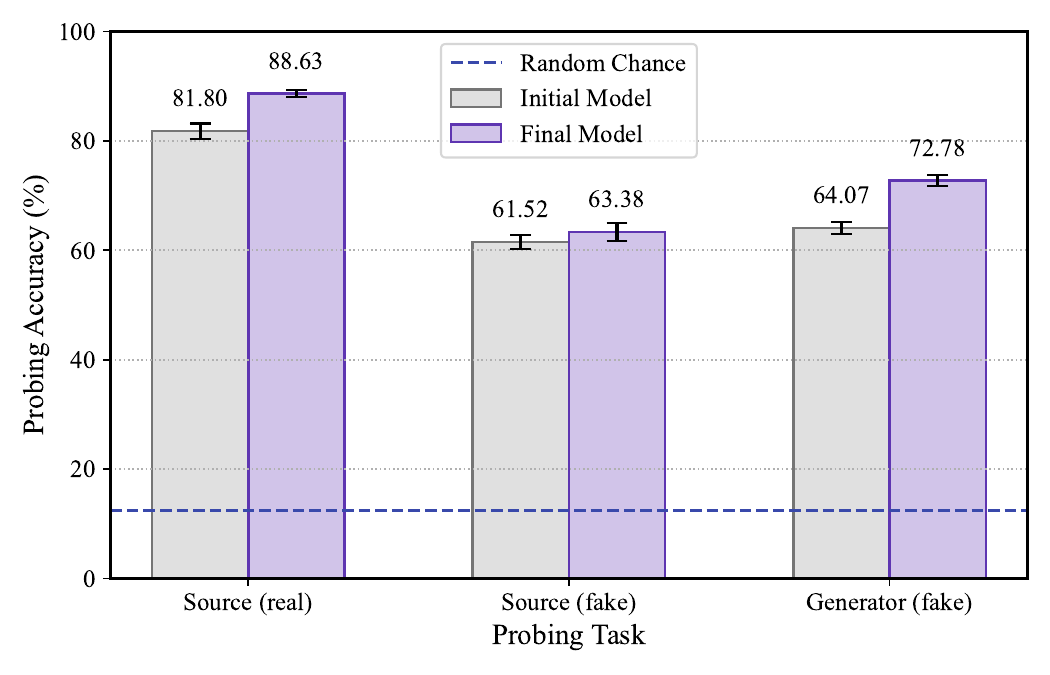}
    \caption{\textbf{Linear probing accuracy of model embeddings.} We compare the initial pre-trained model against our final DOSS-trained model.}
    \label{fig:probing}
\end{figure}

The quantitative probing results, shown in Figure~\ref{fig:probing}, reveal three key findings. \textbf{First}, the Initial Model (pre-trained backbone) already exhibits high probing accuracy for both source and generator identification, far exceeding the random chance baseline. This confirms that self-supervised pre-training provides a powerful representation that already contains rich information about audio attributes. \textbf{Second}, the training actively sharpens this representation. The final model shows a clear performance increase in identifying both real-world sources and, critically, generator artifacts. This verifies our hypothesis that the model learns and potentially leverages these attributes during binary classification.
\textbf{Finally}, the results reveal two informative asymmetries. We observe that probing accuracy for real sources is significantly higher than that for fake sources. This suggests that the generation process might partially obscure the original source information. Furthermore, on the fake data, the model is consistently better at identifying the generator than the underlying source, implying that it could prioritize the generator's artifact as the most dominant signal.

\begin{figure*}[t!]
    \centering
    \begin{subfigure}[b]{0.48\textwidth}
        \centering
        \includegraphics[width=0.8\linewidth]{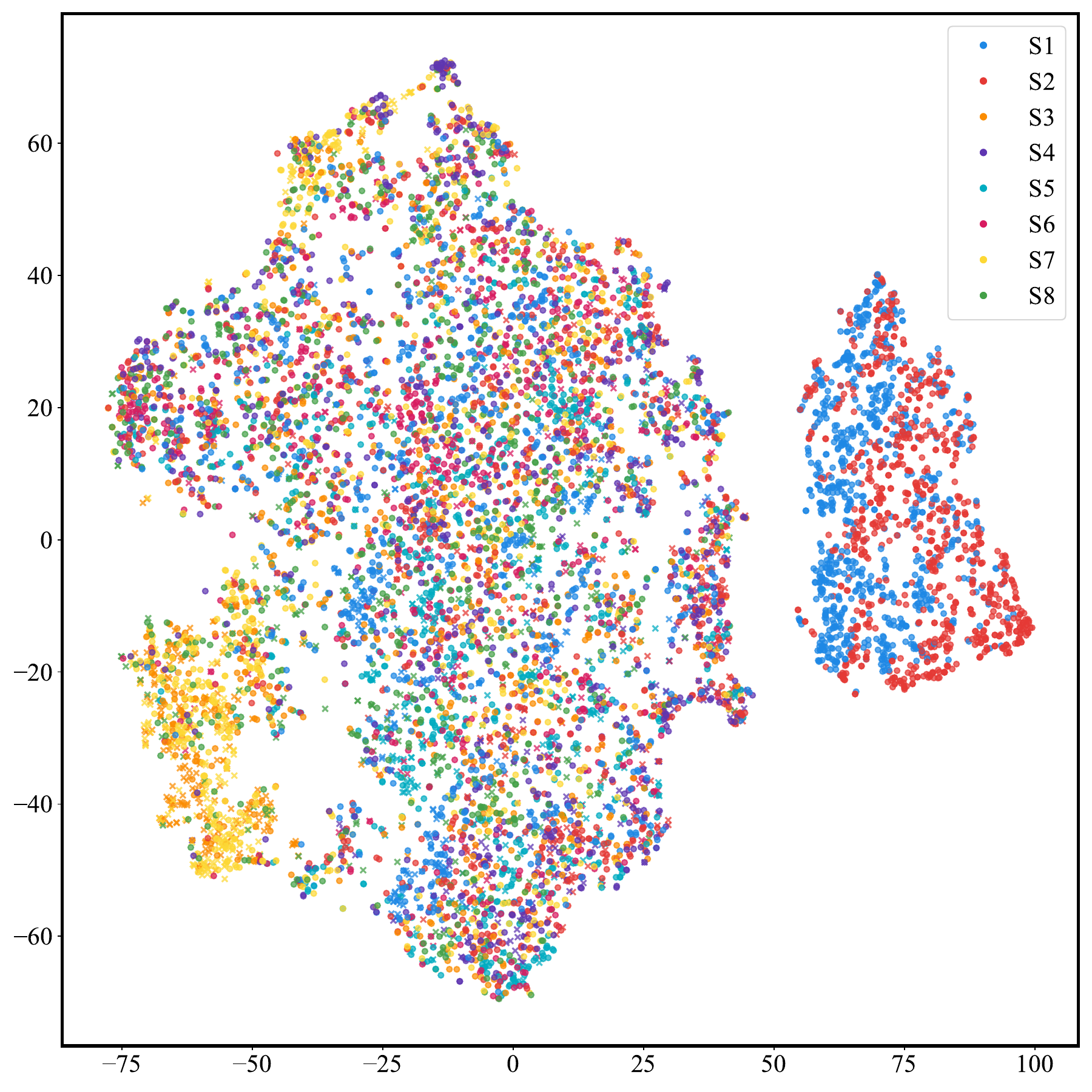}
        \caption{Initial Model: Embeddings colored by Source ID}
        \label{fig:tsne_a}
    \end{subfigure}
    \hfill
    \begin{subfigure}[b]{0.48\textwidth}
        \centering
        \includegraphics[width=0.8\linewidth]{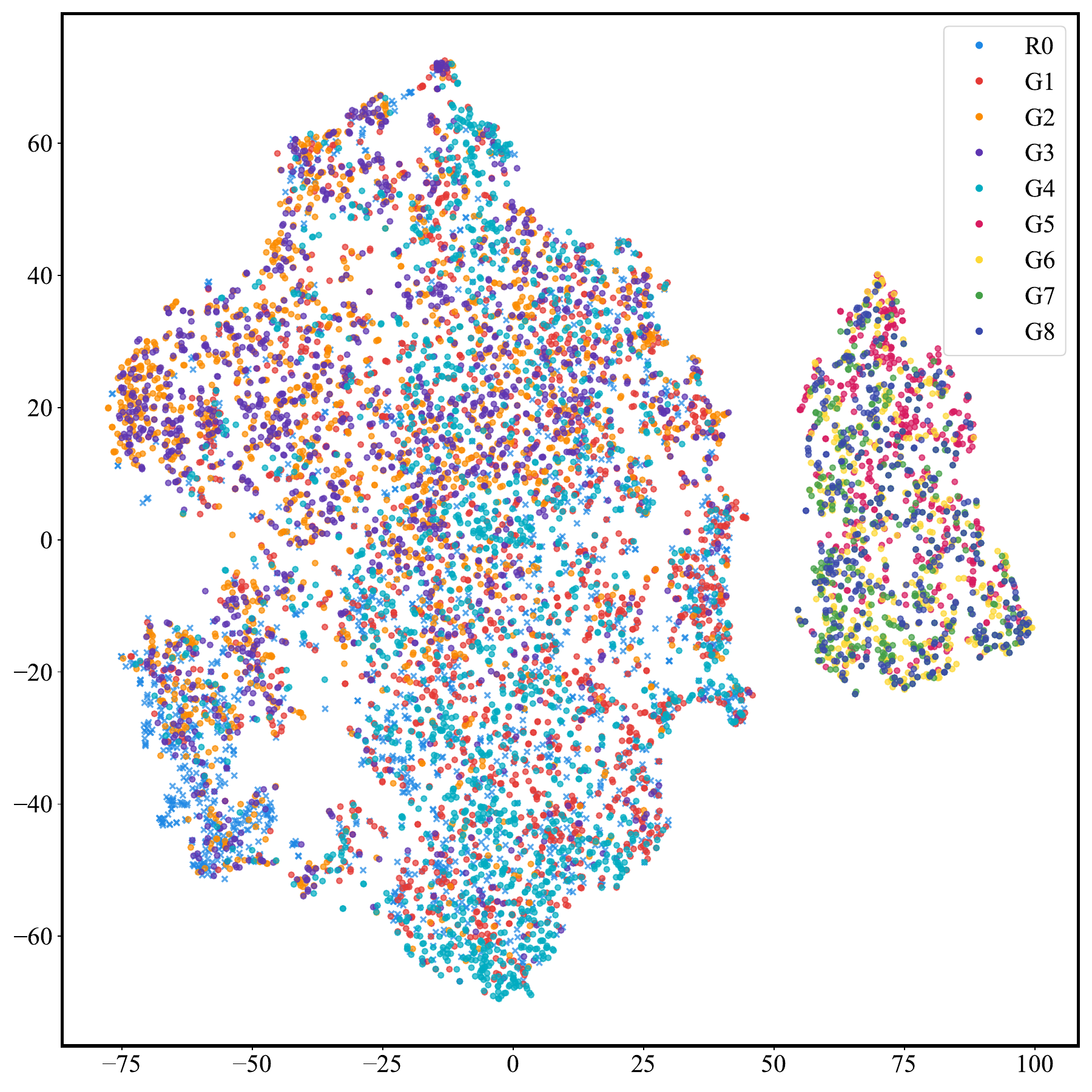}
        \caption{Initial Model: Embeddings colored by Generator ID}
        \label{fig:tsne_b}
    \end{subfigure}

    \begin{subfigure}[b]{0.48\textwidth}
        \centering
        \includegraphics[width=0.8\linewidth]{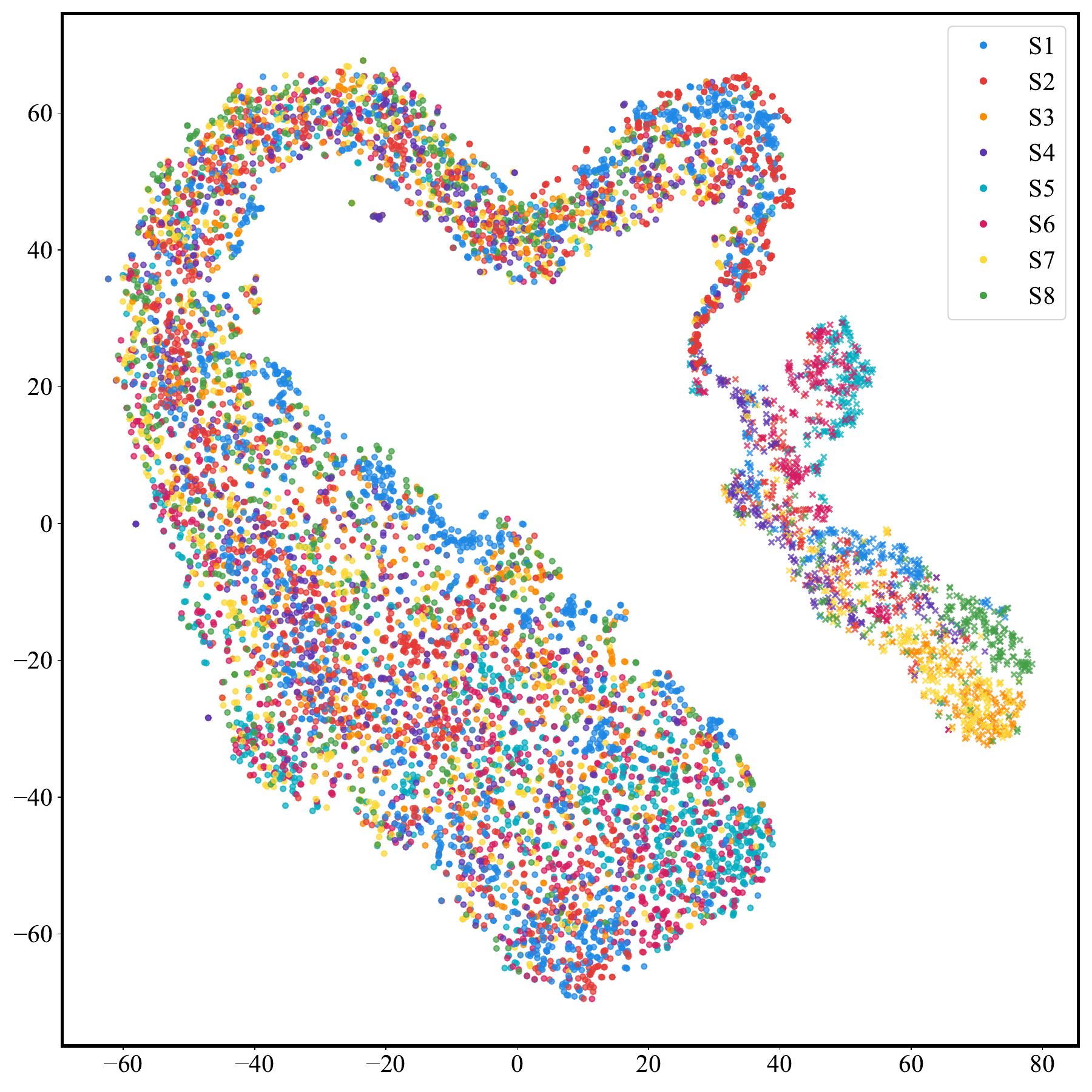}
        \caption{Final Model: Embeddings colored by Source ID}
        \label{fig:tsne_c}
    \end{subfigure}
    \hfill
    \begin{subfigure}[b]{0.48\textwidth}
        \centering
        \includegraphics[width=0.8\linewidth]{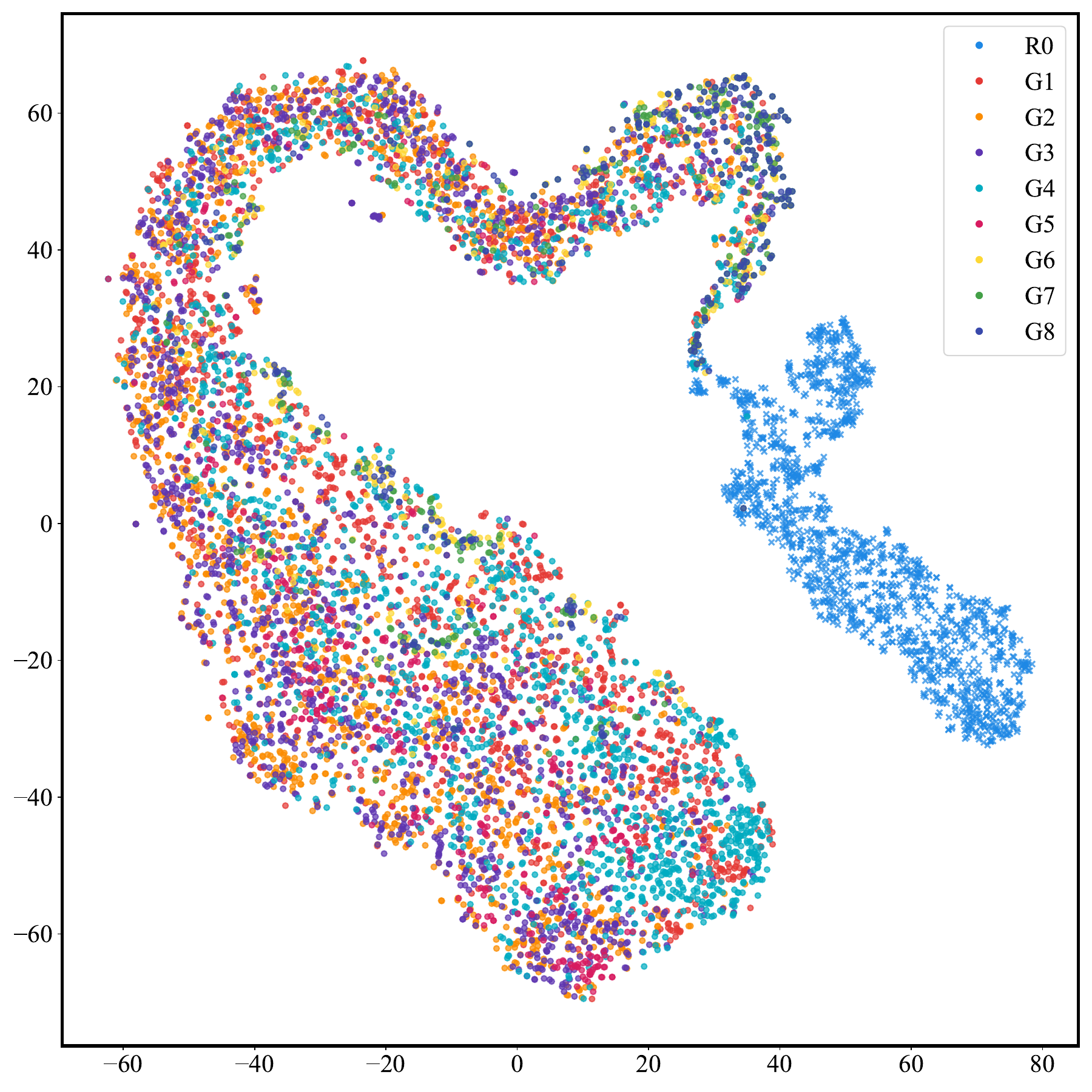}
        \caption{Final Model: Embeddings colored by Generator ID}
        \label{fig:tsne_d}
    \end{subfigure}

    \caption{\textbf{t-SNE visualization of the initial and final model's latent space.} 
             (a-b) show the initial pre-trained model's embeddings. 
             (c-d) show the final DOSS-trained model's embeddings. 
             Across all subplots, crosses represent the real class and circles represent the fake class. In (b) and (d), R0 represents the real class.
            }
    \label{fig:tsne} 
\end{figure*}

The t-SNE visualizations in Figure~\ref{fig:tsne} confirm our quantitative probing results. Figure~\ref{fig:tsne_a} and~\ref{fig:tsne_b} show the latent space of the initial pre-trained model, where embeddings already form nascent clusters based on source and generator. However, the real and fake samples are highly overlapping and difficult to distinguish.
In contrast, Figure~\ref{fig:tsne_c} and~\ref{fig:tsne_d} show that the final DOSS-trained model's latent space is highly structured. While the source and generator clusters are preserved, the real and fake classes are now well-separated. This visual evidence directly supports the high probing accuracies reported in Figure~\ref{fig:probing}. It confirms that the model learns and refines structured representations for these attributes, which it may leverage during the binary classification task.

\end{document}